\documentclass{aa}
\usepackage{epsfig,times}

\def\micron{\,$\mu$m}
\makeatletter
\def\@cite#1#2{(#1\if@tempswa , #2\fi)}
\def\@citex[#1]#2{\if@filesw\immediate\write\@auxout{\string\citation{#2}}\fi
  \def\@citea{}\@cite{\@for\@citeb:=#2\do
    {\@citea\def\@citea{;\penalty\@m\ }\@ifundefined
       {b@\@citeb}{{\bf ?}\@warning
       {Citation `\@citeb' on page \thepage \space undefined}}%
\hbox{\csname b@\@citeb\endcsname}}}{#1}}
\makeatother

\title{Near-infrared detection and optical follow-up of the GRB990705  
afterglow\thanks{Based on observations collected at the European Southern
Observatory, La Silla and Paranal, Chile}$^{,}$\thanks{Based on
observations collected with NOAO facilities}}

\author{N. Masetti\inst{1}
\and E. Palazzi\inst{1}
\and E. Pian\inst{1}
\and L.K. Hunt\inst{2}
\and M. M\'endez\inst{3,4}
\and F. Frontera\inst{1,5}
\and L. Amati\inst{1}
\and P.M. Vreeswijk\inst{3}
\and E. Rol\inst{3}
\and T.J. Galama\inst{3}
\and J. van Paradijs\inst{3,6}$^{,}$\thanks{Deceased on November 2, 1999}
\and L.A. Antonelli\inst{7}
\and L. Nicastro\inst{8}
\and M. Feroci\inst{9}
\and G. Marconi\inst{7}
\and L. Piro\inst{9}
\and E. Costa\inst{9}
\and C. Kouveliotou\inst{10,11}
\and A.J. Castro-Tirado\inst{12,13}
\and R. Falomo\inst{14}
\and T. Augusteijn\inst{15}
\and H. B\"ohnhardt\inst{15}
\and C. Lidman\inst{15}
\and L. Vanzi\inst{15}
\and K.M. Merrill\inst{16}
\and C.D. Kaminsky\inst{17}
\and M. van der Klis\inst{3}
\and M.H.M. Heemskerk\inst{3}
\and F. van der Hooft\inst{3}
\and E. Kuulkers\inst{18,19}
\and H. Pedersen\inst{20}
\and S. Benetti\inst{21}
}

\institute{Istituto Tecnologie e Studio Radiazioni Extraterrestri, CNR,
Via Gobetti 101, I-40129 Bologna, Italy
\and
Centro per l'Astronomia Infrarossa e lo Studio del Mezzo Interstellare,
CNR, Largo E. Fermi 5, I-50125 Firenze, Italy
\and
Astronomical Institute, University of Amsterdam, Kruislaan 403, NL-1098 SJ
Amsterdam, The Netherlands
\and
Facultad de Ciencias Astron\'omicas y Geof\'{\i}sicas, Universidad
Nacional de La Plata, Paseo del Bosque S/N, 1900 La Plata, Argentina
\and
Dipartimento di Fisica, Universit\`a di Ferrara, Via Paradiso 12, I-44100
Ferrara, Italy
\and
Physics Department, University of Alabama in Huntsville, Huntsville,
AL 35899, USA
\and
Osservatorio Astronomico di Roma, Via Frascati 33, I-00040 Monteporzio
Catone, Italy
\and
Istituto di Fisica Cosmica con Applicazioni all'Informatica, CNR, Via U.
La Malfa 153, I-90146 Palermo, Italy
\and
Istituto di Astrofisica Spaziale CNR, Via del Fosso del Cavaliere,
I-00131 Rome, Italy
\and
Universities Space Research Association, Huntsville, AL 35800, USA
\and
NASA Marshall Space Flight Center, ES-84, Huntsville, AL 35812, USA
\and
Instituto de Astrof\'{\i}sica de Andaluc\'{\i}a (IAA-CSIC), P.O. Box
03004, E-18080, Granada, Spain
\and
Laboratorio de Astrof\'{\i}sica Espacial y F\'{\i}sica Fundamental, INTA,
P.O. Box 50737, E-28080, Madrid, Spain
\and
Osservatorio Astronomico di Padova, Vicolo dell'Osservatorio 5, I-35122
Padua, Italy 
\and
European Southern Observatory, Casilla 19000, Santiago, Chile
\and
National Optical Astronomy Observatories, 950 North Cherry Avenue, 
P.O. Box 26732, Tucson, AZ 85726, USA
\and
Center for Astrophysical Research in Astronomy, Yerkes Observatory,
Univ. of Chicago, 373 W. Geneva St, Williams Bay, WI 53191, USA
\and
Space Research Organization Netherlands, Sorbonnelaan 2, NL-3584 CA
Utrecht, The Netherlands
\and
Astronomical Institute, Utrecht University, P.O. Box 80000, NL-3507 TA
Utrecht, The Netherlands
\and
Copenhagen University Observatory, Juliane Maries Vej 30, DK-2100
Copenhagen, Denmark
\and
Telescopio Nazionale Galileo, Centro Galileo Galilei, Calle Alvarez de
Abreu 70, E-38700 Santa Cruz de La Palma, Canary Islands, Spain
}

\offprints{Nicola Masetti, masetti@tesre.bo.cnr.it}
\date{Received July 30, 1999; Accepted \dots}
\thesaurus{03(13.07.1, 02.18.5, 11.07.1)}

\begin{document}

\maketitle
\markboth{N. Masetti et al.: Detection and follow-up of the GRB990705 
afterglow}{}

\begin{abstract}

Optical and near-infrared observations of the GRB990705 error box were
carried out with ESO telescopes at La Silla and Paranal in Chile and with
the NOAO SPIREX 0.6-meter telescope in Antarctica. 
We detected the counterpart of this GRB in the near-infrared $H$ band
and optical $V$ band. The power-law decline of the near-infrared
lightcurve is rather steep with a decay index $\alpha$ $\simeq$ 1.7 in
the first hours, and a possible steepening after one day. Broadband
spectral analysis of
the optical/near-infrared afterglow suggests that this GRB took place in a
high density environment. A deep optical image obtained
at Antu (VLT-UT1) about 5 days after the GRB trigger shows at the position
of the transient an extended object which might be the host
galaxy of GRB990705. 


\end{abstract}

\section{Introduction}

Multiwavelength observations of Gamma--Ray Burst (GRB) afterglows are
of crucial importance for understanding and constraining the active
emission mechanisms (Wijers et al. 1997; Galama et al. 1998, Wijers \&
Galama 1999, Masetti et al. 1999). Optical and near-infrared (NIR) data
carry the richest and most detailed information.  In particular, since the
GRB counterparts might heavily suffer from dust obscuration within the
host galaxy, the NIR data, less affected by this extinction, are more
effective than the optical ones for the study of the counterpart itself
and of the circumburst medium, and, ultimately, in determining the nature
of the GRB progenitors (see e.g. Dai \& Lu 1999).

GRB990705 (Celidonio et al. 1999) was detected by the Gamma-Ray Burst
Monitor (GRBM; Frontera et al. 1997, Amati et al. 1997, Feroci et al.
1997) onboard {\it BeppoSAX} (Boella et al. 1997) on 1999 July 5.66765 UT
and promptly localized with a 3$\arcmin$ accuracy by Unit 2 of the 
{\it BeppoSAX} Wide Field Cameras (WFC; Jager et al. 1997). This GRB
lasted about 45 s in the GRBM 40--700 keV band, in which it reached a
$\gamma$--ray peak flux of (3.7 $\pm$ 0.1)$\times$10$^{-6}$ erg cm$^{-2}$ 
s$^{-1}$ and showed a complex and multi-peaked structure.
The WFC (2--26 keV) data indicate that GRB990705 had a similar
duration and lightcurve in the X--rays, and that it displayed very
bright X--ray emission with a peak intensity of about 4 Crab.
A detailed presentation and description of the prompt event will be given
by Amati et al. (in preparation).

A BeppoSAX X--ray follow-up of GRB990705 started 11 hours after the GRBM
trigger (Gandolfi 1999; Amati et al. 1999).
The detection of this GRB by {\it Ulysses} (Hurley \& Feroci 1999) and
{\it NEAR} (Hurley et al. 1999) determined two annuli intersecting the
{\it BeppoSAX} WFC error circle. This allowed the reduction of the error
box to $\sim$3.5 square arcmin.
Radio observations carried out with ATCA (Subrahmanyan et al. 1999)
detected three radio sources in the WFC error circle.
However, none of them lies inside the intersection of the
{\it Ulysses}, {\it NEAR} and {\it BeppoSAX} error boxes (Hurley et al.
1999).

Optical and near-infrared (NIR) observations were immediately activated at
telescopes in the southern hemisphere to search for a
counterpart at these wavelengths. The early imaging of the 3$\arcmin$
radius WFC error circle at the ESO-NTT with the SOFI camera allowed us to
detect a bright NIR transient (Palazzi et al. 1999) inside the {\it
Ulysses}, {\it NEAR} and {\it BeppoSAX} error boxes intersection.

In this paper we report on the discovery and follow-up
observations of the NIR and Optical Transients (NIRT and OT, respectively)
associated with GRB990705.
In Sect. 2 we describe the data acquisition and reduction, while in Sect.
3 we report the results, which are then discussed in Sect. 4.

\begin{figure*}
\begin{center}
\epsfig{figure=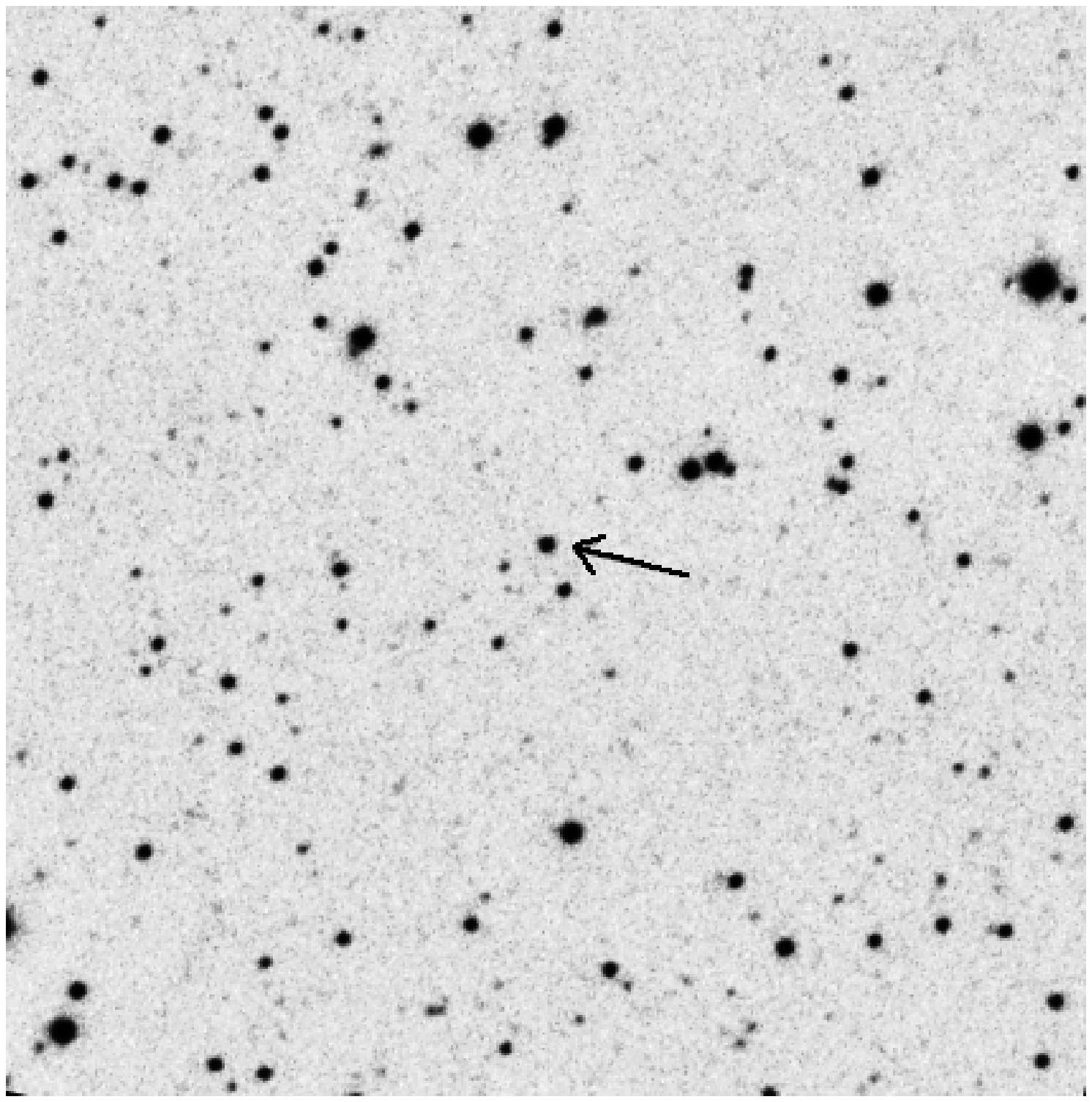,width=8.9cm}
\epsfig{figure=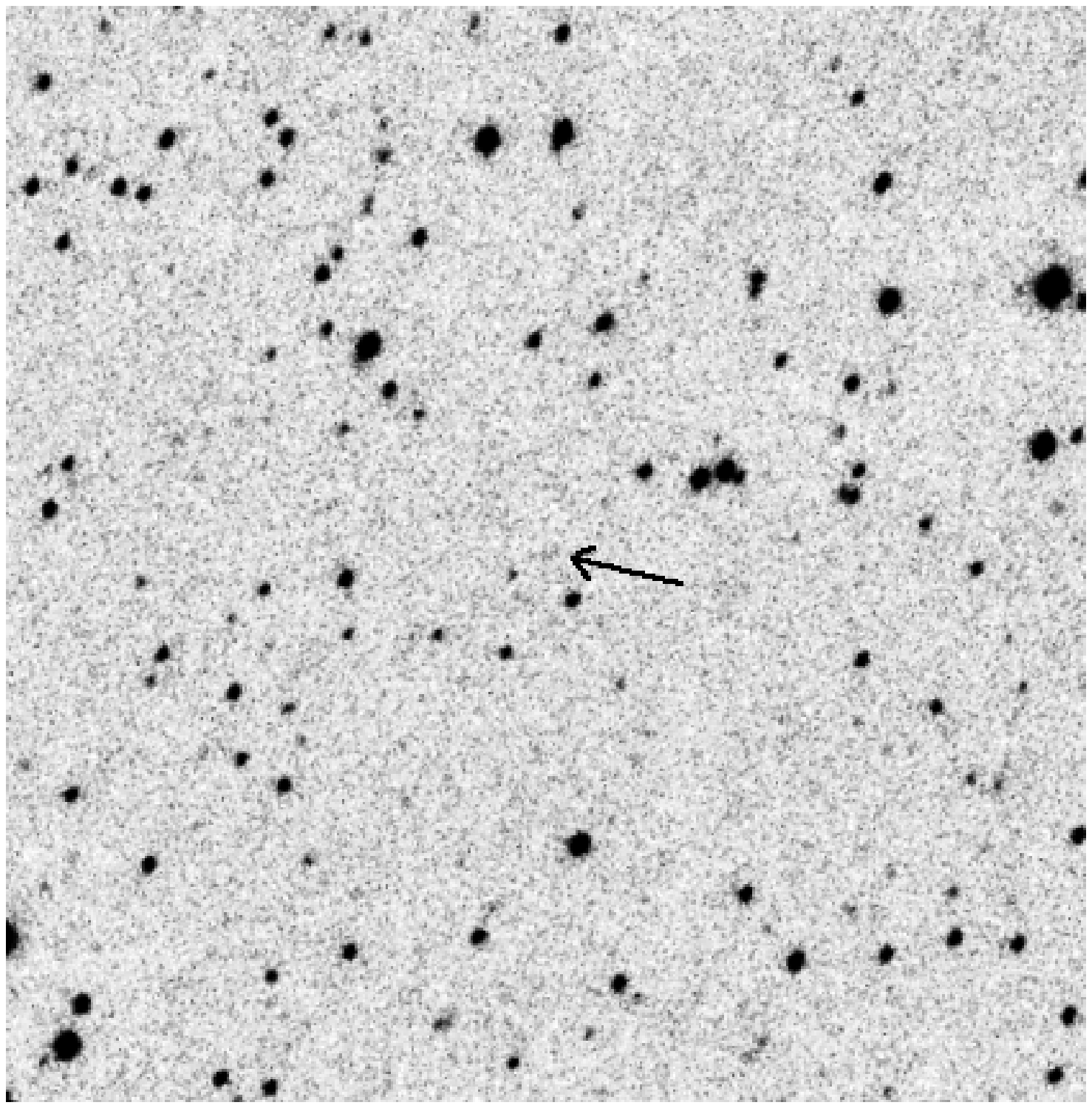,width=8.9cm}
\end{center}
\caption[]{The infrared $H$-band counterpart of GRB990705 (marked with the
arrow) is clearly seen in the NTT+SOFI 1999 July 5.9 summed 20-min. image
({\it left}), while it is barely visible in the 1999
July 6.9 summed 10-min. image ({\it right}) acquired with the same
instrumentation. North is at top, East is to the left; the field size is
about 2$\arcmin$ $\times$ 2$\arcmin$}
\end{figure*}

\section{Observations and data reduction}

\subsection{Near-infrared data}

The NIR imaging started 6.6 hours after the high-energy event:
$H$-band images were acquired on 1999 July 5.9, 6.4 and 6.9 at La Silla
(Chile) with the 3.58-meter ESO-NTT plus SOFI (see the observation log in
Table 1).
The camera is equipped with a Hawaii 1024$\times$1024 pixel
HgCdTe detector, with a plate scale of 0$\farcs$29 pixel$^{-1}$ and a
field of view of roughly 4$\farcm$9$\times$4$\farcm$9.
Images are composed of a number of elementary coadded
frames acquired by dithering the telescope by several arcsecs every 60~s.

Reduction of the images was performed with IRAF and the STSDAS
packages\footnote{IRAF is the Image Analysis and Reduction Facility
made available to the astronomical community by the National Optical
Astronomy Observatories, which are operated by AURA, Inc., under
contract with the U.S. National Science Foundation.
STSDAS is distributed by the Space
Telescope Science Institute, which is operated by the Association of
Universities for Research in Astronomy (AURA), Inc., under NASA contract
NAS 5--26555.}. Each image was reduced by first subtracting a mean sky,
obtained from frames just before and after the source image.
Then, a differential dome flatfield correction was applied, and the
frames were registered to fractional pixels and combined.
Before frames were used for sky subtraction, stars in them
were eliminated by a background interpolation algorithm ({\it imedit})
combined with an automatic ``star finder'' ({\it daofind}).

We calibrated the photometry with stars selected from the
NICMOS Standard List (Persson et al. 1998).
The stars were observed in five positions on the detector,
and were reduced in the same way as the source observations.
Formal photometric accuracy based only on
the standard star observations is typically better than 3\%.
The source photometry was corrected for atmospheric extinction using the
mean ESO $H$-band extinction coefficient of 0.06 (Engels et al. 1981).

$L$-band (3.205-3.823 \micron) observations were also carried out in
Antarctica on July 7.6 with the SPIREX 0.6-meter telescope plus the NOAO
ABU IR camera with a 0$\farcs$6 pixel$^{-1}$ plate scale. ABU houses a
1024$\times$1024 pixel ALADDIN InSb array operating at 36 K using a closed
cycle helium refrigeration system. 

Forty images (eight 5-image cross patterns - center and the cardinal
points - with 30$\arcsec$ separation) of 3-min duration each (12 coadded
15-sec integrations) were obtained for
a total on-source time of 120 minutes. Images were then sky-subtracted
using sky frames generated from the running median of 6 neighbors,
divided by the flatfield, and spatially registered using stars within each
field. The 40 images were shifted and median filtered into the 120-min
composite.
Star HR 2015 (McGregor 1994) was employed as an $L$-band standard to
zero-point calibrate the GRB990705 field.

\subsection{Optical data}

Optical imaging of the GRB990705 error box was obtained at Paranal 
(Chile) with the 8.2-meter ESO VLT-UT1 (``Antu") plus FORS1 (detector
scale: 0$\farcs$2 pixel$^{-1}$; field of view:
6$\farcm$8$\times$6$\farcm$8) on 1999 July 6.4, 8.4 and 10.4 in the $V$
band, and at La Silla (Chile) with the 2.2-meter
MPG/ESO telescope plus WFI (8 CCD mosaic -- detector scale: 0$\farcs$238
pixel$^{-1}$; field of view: 34$\arcmin$$\times$33$\arcmin$) on
1999 July 6.4 ($B$ band) and July 7.4 ($V$ band).
The complete log of the optical observations is reported in Table 1.

\begin{table*}
\caption[]{Journal of the NIR and optical observations of the GRB990705 error
box}
\begin{tabular}{ccccccl}
\noalign{\smallskip}
\hline
\noalign{\smallskip}
Day of 1999 (UT) & Telescope & Filter & Exp. time & Seeing & 
Magnitude$^1$ & \\
at exposure start & & & (minutes) & (arcsecs) &  & \\
\noalign{\smallskip}
\hline
\noalign{\smallskip}
\multicolumn{7}{c}{Near-Infrared} \\
\noalign{\smallskip}
\hline
\noalign{\smallskip}
\multicolumn{1}{l}{Jul 5.945} & NTT & $H$ & 20$\times$1 & 1.0 &
16.57 $\pm$ 0.05$^2$ &\\
\multicolumn{1}{l}{Jul 6.416} & NTT & $H$ & 8$\times$1& 1.2 &
18.38 $\pm$ 0.05 &\\
\multicolumn{1}{l}{Jul 6.955} & NTT & $H$ & 10$\times$1 & 1.1 &
$>$19.9$^2$ &\\
\multicolumn{1}{l}{Jul 7.556} & SPIREX & $L$ & 40$\times$3 & $\sim$1.8 & 
$>$13.9 &\\
\noalign{\smallskip}
\hline
\noalign{\smallskip}
\multicolumn{7}{c}{Optical} \\
\noalign{\smallskip}
\hline
\noalign{\smallskip}
\multicolumn{1}{l}{Jul 6.400} & Antu & $V$ & 5$\times$2 & 2.5 &
22.0 $\pm$ 0.2 &\\
\multicolumn{1}{l}{Jul 6.444} & 2.2m & $B$ & 7.5 & 1.8 & 
$>$21.9 &\\
\multicolumn{1}{l}{Jul 7.432} & 2.2m & $V$ & 2$\times$10 & 2.0 & 
$>$22.3 &\\
\multicolumn{1}{l}{Jul 8.425} & Antu & $V$ & 1 & 1.2 & 
$>$23.0 &\\
\multicolumn{1}{l}{Jul 10.401} & Antu & $V$ & 6$\times$5 & 0.9 &
23.99 $\pm$ 0.07 & (pointlike object, likely unrelated to the OT)\\
 & & & & & 23.8 $\pm$ 0.2 & (extended object)\\
 & & & & & 23.14 $\pm$ 0.15 & (their integrated magnitude)\\
\noalign{\smallskip}
\hline
\noalign{\smallskip}
\multicolumn{7}{l}{$^1$Magnitudes of the GRB counterpart, not corrected
for interstellar absorption}\\
\multicolumn{7}{l}{$^2$Uncertainties of the magnitudes are at
1$\sigma$ confidence level; lower limits at 3$\sigma$}\\
\noalign{\smallskip}
\hline
\end{tabular}
\end{table*}

Images were debiased and flat-fielded with the standard cleaning procedure;
each set of $V$ frames of July 6, 7, and 10 was then co-added to increase  
the signal-to-noise ratio.
We then chose, when applicable, PSF-fitting photometry as the measurement
technique for the magnitude of point-like objects because the field is
quite crowded (especially in the case of deep images) being located in the
outskirts of the Large Magellanic Cloud (LMC). 
Photometry was performed on the images
using the DAOPHOT II data analysis package PSF-fitting algorithm (Stetson
1987) within {\sl MIDAS}.

In order to calibrate the images to the Johnson-Kron-Cousins
photometric system, we acquired on July 7 a $B$ frame of part of Selected
Area 95 with the 2.2-meter telescope and on July 10 $V$ frames of the
PG 1323$-$086, PG 2213$-$006 and Mark A sequences (Landolt 1992) with 
Antu; we adopted as airmass extinction coefficients 0.11 for the $V$ and
0.25 for the $B$.

With this photometric calibration, for comparison, the
USNO-A1.0 star U0150\_02651600, with
coordinates (J2000) $\alpha$ = 5$^{\rm h}$ 09$^{\rm m}$ 42$\fs$59,
$\delta$ = $-$72$^{\circ}$ 07$\arcmin$ 41$\farcs$2, has $B$ = 18.72 and
$V$ = 17.64.
Unfortunately, the $B$ calibration frames were obtained under poor
photometric conditions (see Table 1) and therefore the uncertainty on the
zero point of the calibration ($\pm$0.25 mag) is by far the main source of
error in the measure of the $B$ magnitude of the USNO star. No
color term was applied in the $V$-band calibration since only $V$
frames were taken on July 10; so, the uncertainty on the $V$ zero
point is $\pm$0.15 mag. These large errors are also due to the high airmass
(larger than 2) affecting our observations.
The $B$ and $V$ magnitude errors quoted in the next section are only
statistical and do not contain any possible zero point offset.

\begin{figure}
\vspace{-0.8cm}
\epsfig{figure=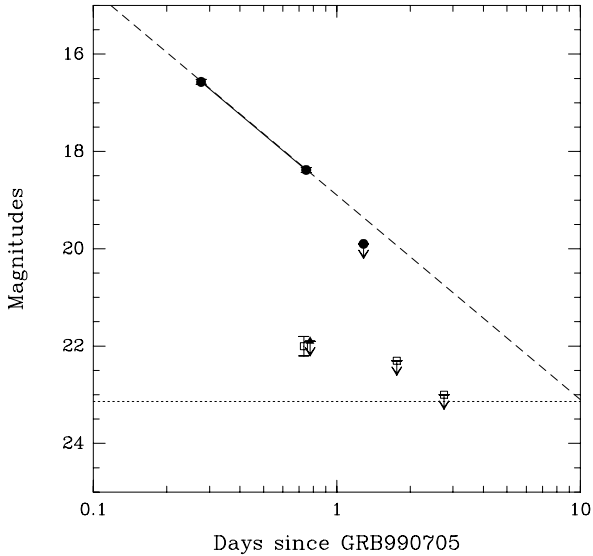,width=9.5cm}
\vspace{-1.3cm}
\caption[]{Optical and NIR lightcurves of the GRB990705 afterglow. 
In abscissa the time elapsed since GRB trigger ($t_{\rm GRB}$: 1999 July
5.66765 UT) is reported. Filled
circles and open squares represent the $H$- and $V$-band measurements (or
upper limits), respectively, and the single $B$-band upper limit is
represented with a filled triangle. Error bars are 1$\sigma$
uncertainties, while upper limits correspond to a 3$\sigma$ confidence
level. The $H$-band decay with index
$\alpha$ = 1.7 computed between the first two epochs is shown
as a solid line, and its extrapolation is indicated with the dashed
line. The dotted horizontal line represents the integrated optical
magnitude ($V$ = 23.14) of the extended and point-like objects measured on
1999 July 10 at the position of the OT} 
\end{figure}

We evaluated the Galactic hydrogen column density in the direction of
GRB990705 using the NRAO maps by Dickey \& Lockman (1990), from which we
obtained $N_{\rm H}$ = 0.72$\times$10$^{21}$ cm$^{-2}$ and, using the
empirical relationship by Predehl \& Schmitt (1995), we computed a
foreground Galactic absorption $A_V$ = 0.40.
This, by applying the law by Rieke \& Lebofsky (1985), corresponds to
$E(B-V)$ = 0.13 and to $E(V-H)$ = 0.33; using the law by Cardelli et al.
(1989) we then derived $A_B$ = 0.53 and $A_H$ = 0.07. The
intrinsic value of $N_{\rm H}$ in that region of the LMC is
less than $\approx$10$^{19}$ cm$^{-2}$ (McGee et al. 1983); therefore, the
reddening induced by the LMC on the NIRT/OT is practically negligible.

\begin{figure*}
\epsfig{figure=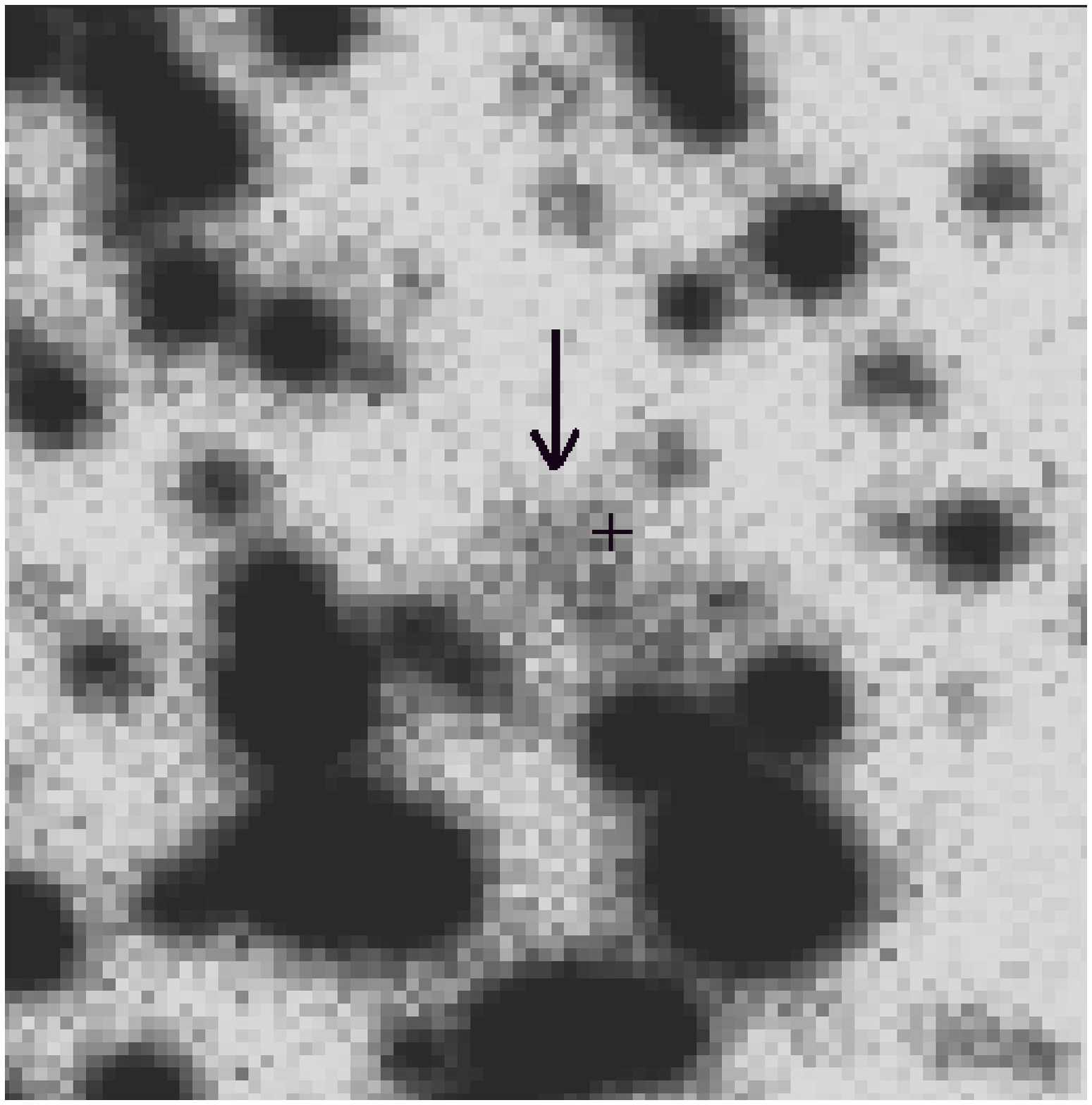,height=9.0cm}
\epsfig{figure=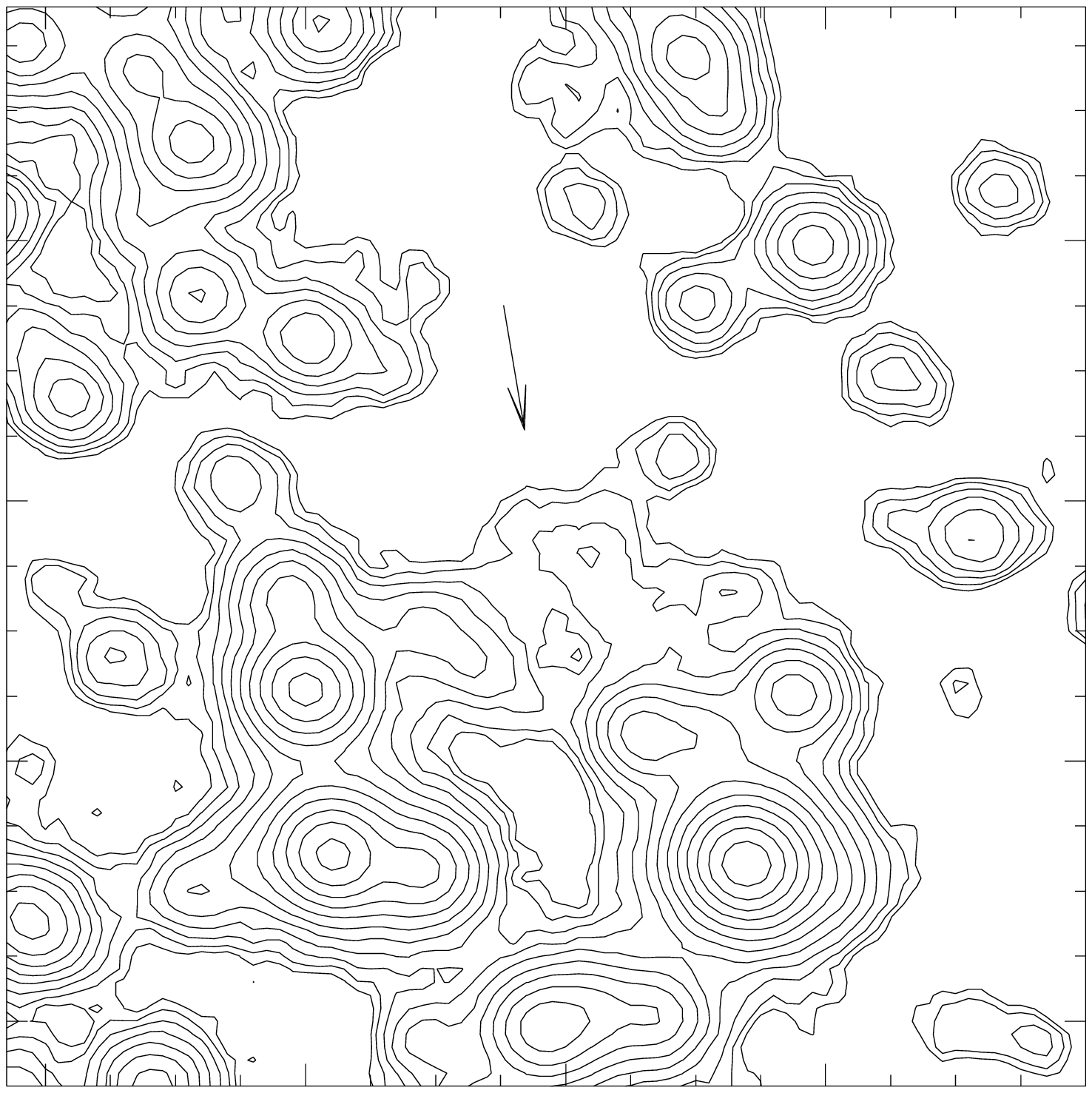,height=9.0cm}
\caption[]{({\it Left panel}) VLT-Antu 30-min $V$ image of the putative
GRB990705 host galaxy
(indicated with the arrow) acquired on 1999 July 10. North is at top, East is
to the left; the field size is about 10$\arcsec$ $\times$ 10$\arcsec$.
The cross shows the position of the NIRT.
({\it Right panel}) Isophotal contour plot of the same region
obtained from the image after slight gaussian filtering. The spacing
between isophotes is 0.5 mag (lowest level is 5 mag below the sky
brightness). The arrow marks the extended irregular structure (see text).
}
\end{figure*}

\section{Results}

The summed 20-min NTT image of July 5.9 (Fig. 1, left panel) shows an
object at a magnitude $H$ = 16.57 $\pm$ 0.05 which in the July 6.4 8-min
image is detected at $H$ = 18.38 $\pm$ 0.05. On July 6.9 the object
magnitude is $H$ $>$ 19.9 at a 3$\sigma$ level (Fig. 1, right panel).
Astrometry done on the first NTT observation using several stars
from the USNO-A1.0 catalogue gives for this fading
source coordinates $\alpha$ = 5$^{\rm h}$ 09$^{\rm m}$ 54$\fs$52,
$\delta$ = $-$72$^{\circ}$ 07$\arcmin$ 53$\farcs$1 (J2000) with a
1-$\sigma$ accuracy of 0$\farcs$3.
This object is inside the intersection of all the mentioned X--ray error
boxes, and almost at the center of the {\it BeppoSAX} WFC error circle.
Moreover, the observed brightness variation and the variability
timescale are similar to those of previously observed optical afterglows.
This leads us to conclude that it is the NIR afterglow of GRB990705
(we can exclude a LMC microlensing event since these
phenomena show a completely different behaviour: see Sackett 1999 and
references therein). 

No object is detected in the $L$-band July 7.6 composite image. An upper
limit $L$ $>$ 13.9 with a 3$\sigma$ significance within a 3
pixel radius aperture at the location corresponding to the $H$-band
detection is measured.

Assuming a temporal power-law decay $F \propto t^{-\alpha}$ between the two
$H$-band detections, we find that $\alpha$ = 1.68 $\pm$ 0.10. 
Including the $H$-band upper limit and fitting a power law the decay
exponent is $\alpha$ = 1.84 $\pm$ 0.05, but the fit is not acceptable
($\chi^2_{\nu}$ = 16.8). In the following we will thus consider $\alpha$
= 1.7 as the decay index of the early part of the afterglow while,
$\sim$1 day after the GRB, the transient has probably started a faster
decay with a power-law index $\alpha$$\arcmin$ $>$ 2.6, based on the
second $H$-band detection and the $H$-band upper limit. 
We note however that the paucity of the data makes difficult to precisely
locate the epoch at which the decay slope has changed.

Antu $V$-band observations, albeit with lower significance, due to the
faintness of the object and to poorer weather conditions (see Table 1),
are consistent with the NIR decay: indeed a fading optical object at a
position consistent with that of the NIRT is detected. On July 6.4 this
object was at $V$ = 22.0 $\pm$ 0.2, while two days later it was below the
limiting magnitude of the frame ($V$ = 23.0, 3$\sigma$ level). These
values indicate a power-law decay with an index $\alpha >$ 1.0, consistent
with the NIR observations.

The object is not seen in the 2.2-meter $B$ frame of July 6 down to a
limiting magnitude of $B$ = 21.9 and in the $V$ frame acquired with the
same telescope on July 7 down to $V$ = 22.3 (both values have a 3$\sigma$
significance).

The NIR and optical photometry is reported in Fig. 2, where the $H$-band
early decay is also modeled as a power law.

Inspection of the Antu summed 30-min image of July 10 (see Fig. 3,
left panel) shows an irregular extended object at the
location of the NIRT and of the OT. We estimate that the magnitude of
this object ($\sim$ 2$\farcs$4$\times$0$\farcs$8 in size)
is $V$ = 23.8 $\pm$ 0.2. Due to the poor resolution we are not able
to exclude that this feature (or a portion of it) is due to the
contribution of many unresolved sources in the LMC and/or background
faint sources. Our photometric analysis of the field, however,
reveals only one possible point-like object at $V$ = 23.99 $\pm$ 0.07 in
this area southward of the NIRT.

After comparison of positions of field stars in Antu and NTT images,
the position of the point-like object is not completely
consistent with that of the NIRT, being $\sim$1$\farcs$2 (i.e. more than
4$\sigma$) away from it (in the first Antu image, where the OT is
detected, the extremely bad seeing hampers a significant positional
comparison).
Assuming that the point-like source is the transient, it is hard to
explain this offset as the result of a possible contribution of the
extended source in the first $H$-band observation, because if this effect
were present, the centroid of the NIRT+galaxy blend would be expected
to be closer to the fuzziness center than that observed in the first
NTT observation.
Moreover, the $V$-band temporal decay would be much slower than that in
the NIR ($\alpha_V$ $\sim$ 0.9). Therefore we suggest that the point-like
object seen on July 10 is unrelated to the GRB, and might rather be a
structure of the host galaxy, or possibly a foreground star.

The transient, as observed about 0.8 days after the GRB trigger, is fairly
red. In order to evaluate its color index between the $V$ and $H$ bands,
we assumed in the $V$ band a temporal decay similar to that observed in 
the $H$ band and computed the $V$ magnitude at the epoch of the
second $H$-band observation (July 6.416). Then, we subtracted from this
$V$ magnitude the contribution of the nebulosity and of the neighboring
point-like object (which is blended with the OT in the July 6.4 image) and
from the $H$ magnitude of July 6.4 the upper limit of the third $H$-band
observation ($H$ $>$ 19.9).
Finally, we corrected for the Galactic extinction in the direction of
GRB990705. We find $(V-H)_{\rm OT}$ = 3.5 $\pm$ 0.2 on July 6.416;
if instead the host galaxy contribution is much fainter than
the third epoch upper limit, and therefore negligible in the second
$H$-band measurement, the color index is $(V-H)_{\rm OT}$ = 3.8 $\pm$ 0.2.
Both values imply a spectral slope $\beta$ $\sim$ 2 assuming a
spectral energy distribution $F_\nu \propto \nu^{-\beta}$.
This spectral slope would result in a magnitude $L$ $\sim$ 15.4 at
that epoch, i.e. $\sim$1.5 magnitudes fainter than, and therefore fully
consistent with, the upper limit from the SPIREX observation of more than
one day after.

\section{Discussion: a ``red-heat" GRB afterglow}

The afterglow of GRB990705 is an unprecedented case of a GRB
counterpart first clearly detected in the NIR band.
The detection of a possible underlying galaxy might support the
extragalactic nature of this GRB although we cannot completely rule
out an association with the LMC.

If the extended emission detected is the host galaxy of GRB990705 it
seems to have a rather knotty and irregular shape since no
regular pattern of increasing surface brightness is observed in this
structure (see Fig. 3, right panel). The present data therefore
suggest that the host of GRB990705 is an irregular (possibly
starburst) galaxy as was proposed in other cases of GRB hosts
(see e.g. Sahu et al. 1997 for the host of GRB970228, and Bloom et
al. 1999 and Fruchter et al. 1999b for the host of GRB990123). 

With $V$ $\sim$ 23.1, and assuming it is an irregular starburst galaxy
with a flat optical spectrum (Fruchter et al. 1999b), we obtain for this
object $R\sim$ 22.8. Using the cumulative surface density
distribution of galaxies in the $R$ band by Hogg et al. (1997),
the probability $P_c$ of a chance coincidence between the NIR/optical
afterglow of GRB990705 and the detected galaxy can be evaluated. We have
2.5$\times$10$^4$ galaxies per square degree with $R\leq$ 22.8; with this
value the probability of finding by chance a galaxy within 3$\sigma$ from
the position of the NIRT is $P_c\sim$ 0.006. This probability suggests the
identification of this object with the host galaxy, although it is not
completely conclusive.

This putative galaxy has an integrated (point-like source plus extended
object) unabsorbed magnitude $V_0$ = 22.74 $\pm$ 0.15, an
extension of $\sim$2 square arcsecs and an irregular shape (see Fig. 3);
therefore, it might be one of the brightest and most extended among the
host galaxies of GRBs with known redshifts (GRB970228: Sahu et al. 1997,
Fruchter et al. 1999a, Djorgovski et al. 1999; GRB970508: Bloom et al.
1998a, Fruchter et al. 1999c; GRB971214: Kulkarni et al. 1998, Odewahn et
al. 1998; GRB980703: Bloom et al. 1998b, Vreeswijk et al. 1999; and
GRB990123: Kulkarni et al. 1999, Bloom et al. 1999, Fruchter et al.
1999b). This might suggest that this object is nearer than the other GRBs.

The decay slope of this afterglow, $\alpha$ = 1.7, is rather
steep, although not as steep as observed for GRB980326 (Groot et al. 1998)
and GRB980519 (Djorgovski et al. 1998, Halpern et al. 1999). 
From the $H$-band light curve decay index we estimate an electron
power-law distribution index $p$ $\sim$ 3 (Sari et al. 1999).

As already outlined in Sect. 3, the $H$ magnitude of the NIRT
on July 6.9 is significantly below the extrapolation of the early decay
(see Fig. 2). This strongly suggests a break in the NIRT $H$-band light
curve at $\sim$1 day after the GRB and a subsequent steepening, similar
to those exhibited by the afterglows of GRB990123 (e.g. Castro-Tirado et
al. 1999) and of GRB990510 (e.g. Stanek et al. 1999).

The break cannot be accounted for by the electron cooling frequency
$\nu_{\rm c}$ moving through the $H$ band since
the expected slope change ($\Delta\alpha$ $\sim$ 0.25; Sari et
al. 1998) would be much smaller than observed ($\Delta\alpha$ $\ga$ 1).

A spherical scenario in which an extremely dense
surrounding medium decelerates the expanding blastwave could also produce
a steepening of the light curve as envisaged by Dai \& Lu (1999).

On the other hand, the steepness of the lightcurve decay might suggest
beamed emission (Sari et al. 1999).
Assuming that a break due to jet spreading occurred in the
$H$-band light curve of the GRB990705 NIRT about one day after the GRB,
the slope $\alpha$$\arcmin$ $>$ 2.6 would be roughly consistent with the
expected value ($\alpha$$\arcmin$ = $p\sim$ 3; Sari et
al. 1999). If we place the break
at the epoch of the second $H$-band measurement, i.e. $\sim$18 hours
after the GRB trigger, assuming a total isotropically-emitted energy
of the ejecta of 10$^{\rm 52}$ erg and a local interstellar medium density
of 1 cm$^{\rm -3}$, we obtain that the angular width of the jet is
$\theta_0 \sim$ 0.1, consistent with expected opening angles of GRB jets
(Sari et al. 1999, Postnov et al. 1999).

This afterglow is also one of the reddest observed so far, with an
optical/NIR color on July 6.4 similar to that of the OT of GRB980329
(Palazzi et al. 1998). 
We note that the NIR/optical spectral slope $\beta$ of the
transient on July 6.4 and the measured index of the temporal decay
$\alpha$ would be inconsistent both with the spherical expansion of a
relativistic blast wave (assumed as a valid approximation of an initially
strongly beamed jet) and with a beamed expansion (Sari et al. 1999).
Under the hypothesis that the optical/NIR spectrum is considerably
reddened by absorption within the host galaxy, we corrected it using the
extinction law of a typical starburst at various redshifts (Calzetti
1997). This approach, which we have adopted also for other GRBs with
encouraging results (Palazzi et al. 1998, Dal Fiume et al. 2000), can be
justified under the assumption that the heavy obscuration of this
afterglow is due to its location in a high density, and probably
star-forming, region.

We find a consistency with the expectation of the model by Sari et al.
(1999) either for a redshift $z \sim$ 2 or for a redshift $z \sim$ 0.1. In
both cases $\nu_{\rm c}$ must be above the NIR frequencies, which is a
reasonable finding given that the optical/NIR spectrum is measured at an
early epoch after the GRB.
Since the host galaxy is bright and rather large in angular size, we tend
to favor the latter redshift estimate. This result has to be taken with
caution, being based on a series of assumptions and on a single color
index, and therefore affected by a large uncertainty.
However, if it were correct, the emitted $\gamma$-ray output, for a
fluence of 7.5$\times$10$^{-5}$ erg cm$^{-2}$ (Amati et al., in
preparation), would
be 1.7$\times$10$^{51}$ erg (assuming a standard Friedmann model
cosmology with $H_0$ = 70 km s$^{-1}$ Mpc$^{-1}$ and $q_0$ = 0.15), in the
range of $\gamma$-ray energies typically measured for GRBs.

\begin{acknowledgements}

M. M\'endez is a fellow of the Consejo Nacional de Investigaciones
Cient\'{\i}ficas y T\'ecnicas de la Rep\'ublica Argentina.
This research was supported in part by the National Science Foundation
under a cooperative agreement with the Center for Astrophysical Research
in Antarctica (CARA), grant number NSF OPP 89-20223. CARA is a National
Science Foundation Science and Technology Center. Principal collaborators
in the SPIREX/ABU project include the University of Chicago, the National
Optical Astronomy Observatories, the Rochester Institute of Technology,
and the University of New South Wales. We owe special thanks for the
SPIREX/ABU data to Ian Gatley, Nigel Sharp, Al Fowler and Harvey Rhody.
We also thank the referee J. Lub for his comments.

\end{acknowledgements}

\end{document}